\documentclass[12pt]{article}
\usepackage{amsmath, amsthm, amssymb, amsfonts, enumerate,bm,url}
\usepackage[figuresright]{rotating}
\usepackage{lscape}
\usepackage{multirow}
\usepackage{natbib}

\begin{document}
\newcommand{\abs}[1]{\left\vert#1\right\vert}
\newcommand{\set}[1]{\left\{#1\right\}}
\newcommand{\eps}{\varepsilon}
\newcommand{\To}{\rightarrow}
\newcommand{\inv}{^{-1}}
\newcommand{\ihat}{\hat{\imath}}
\newcommand{\var}{\mbox{Var}}
\newcommand{\sd}{\mbox{SD}}
\newcommand{\cov}{\mbox{Cov}}
\newcommand{\f}{\frac}
\newcommand{\fI}[1]{\frac{1}{#1}}
\newcommand{\what}[1]{\widehat{#1}}
\newcommand{\hhat}[1]{\what{\what{#1}}}
\newcommand{\wtilde}[1]{\widetilde{#1}}
\newcommand{\bdot}{\bm{\cdot}}
\newcommand{\Th}{\theta}
\newcommand{\qmq}[1]{\quad\mbox{#1}\quad}
\newcommand{\qm}[1]{\quad\mbox{#1}}
\newcommand{\mq}[1]{\mbox{#1}\quad}
\newcommand{\tr}{\mbox{tr}}
\newcommand{\logit}{\mbox{logit}}
\newcommand{\noi}{\noindent}
\newcommand{\bni}{\bigskip\noindent}
\newcommand{\bul}{$\bullet$ }
\newcommand{\bias}{\mbox{bias}}
\newcommand{\conv}{\mbox{conv}}
\newcommand{\spn}{\mbox{span}}
\newcommand{\colspace}{\mbox{colspace}}
\newcommand{\mX}{\mathcal{X}}
\newcommand{\mT}{\mathcal{T}}
\newcommand{\mF}{\mathcal{F}}
\newcommand{\mI}{\mathcal{I}}
\newcommand{\mS}{\mathcal{S}}
\newcommand{\bbR}{\mathbb{R}}
\newcommand{\fweI}{FWE$_I$}
\newcommand{\fweII}{FWE$_{II}$}
\newcommand{\vphi}{\varphi}

\newtheorem{theorem}{Theorem}[section]
\newtheorem{corollary}{Corollary}[section]
\newtheorem{conjecture}{Conjecture}[section]
\newtheorem{proposition}{Proposition}[section]
\newtheorem{lemma}{Lemma}[section]
\newtheorem{definition}{Definition}[section]
\newtheorem{example}{Example}[section]
\newtheorem{remark}{Remark}[section]

\title{{\bf\Large Sequential Tests of Multiple Hypotheses
    Controlling Type~I and II Familywise Error Rates}}

\author{\textsc{Jay Bartroff\footnote{Department of Mathematics, University of Southern California, Los Angeles, California, USA} and Jinlin Song\footnote{Analysis Group, 111 Huntington Avenue,
Tenth Floor, Boston, MA  02199, USA}}}  
\footnotetext{Key words and phrases: Clinical trials, Holm's procedure, Multiple testing, Sequential analysis, Step-down test, Wald approximations.} 
\footnotetext{\textit{Email addresses}: bartroff@usc.edu (J.\ Bartroff), jsong@analysisgroup.com (J.\ Song)} 

\date{}
\maketitle

\abstract{This paper addresses the following general scenario: A scientist wishes to perform a battery of experiments, each generating a sequential stream of data,  to investigate some phenomenon. The scientist would like to control the overall error rate in order to draw statistically-valid conclusions from each experiment, while being as efficient as possible. The between-stream data may differ in distribution and dimension but also may be highly correlated, even duplicated exactly in some cases. Treating each experiment as a hypothesis test and adopting the familywise error rate (FWER) metric,  we give a procedure that sequentially tests each hypothesis while controlling both the type~I and II FWERs regardless of the between-stream correlation, and only requires  arbitrary sequential test statistics that control the error rates for a given stream in isolation. The proposed procedure, which we call the sequential Holm procedure  because of its inspiration from Holm's (1979) seminal fixed-sample  procedure, shows simultaneous savings in expected sample size and less conservative error control relative to fixed sample, sequential Bonferroni, and other recently proposed sequential procedures in a simulation study.} 

\section{Introduction}

This paper addresses the following scenario: A scientist wishes to perform a battery of $k\ge 2$ experiments sequentially in time in order to investigate some phenomenon, resulting in $k$ \emph{data streams}:
\begin{align}
&\mq{Data stream $1$}X_1^{(1)}, X_2^{(1)},\ldots\qm{from Experiment $1$}\nonumber\\
&\mq{Data stream $2$}X_1^{(2)}, X_2^{(2)},\ldots\qm{from Experiment $2$}\label{streams}\\
&\vdots\nonumber\\
&\mq{Data stream $k$}X_1^{(k)}, X_2^{(k)},\ldots\qm{from Experiment $k$.}\nonumber
\end{align} The scientist would like to control the overall error rate of the battery of experiments in order to be able to draw statistically-valid conclusions for each experiment once all experimentation has ceased, but also needs to be as efficient as possible with the finite resources available by ``dropping'' certain experiments (i.e., stopping experimentation) when additional data is no longer needed from that stream to reach a conclusion.  The between-stream data may be very dissimilar in distribution and dimension, but at the same time may be highly correlated, or even duplicated exactly in some cases, since they all are related to some phenomenon.

The preceding scenario occurs in a number of real applications including multiple endpoint (or multi-arm) clinical trials \citep[][Chapter~15]{Jennison00}, multi-channel changepoint detection \citep{Tartakovsky03} and its applications to biosurveillance \citep{Mei10}, genetics and genomics \citep{Dudoit08}, acceptance sampling with multiple criteria \citep{Baillie87}, and financial trading strategies \citep{Romano05b}. If we think of each experiment as a hypothesis test about the  corresponding data stream, then what is needed is a combination of a multiple hypothesis test and a sequential hypothesis test. We point out that our use of the word ``sequential'' here and below refers to the manner of sampling (or equivalently, observation) and differs from the way the word is sometimes used in the literature on fixed-sample multiple testing procedures to describe the stepwise analysis of fixed-sample test statistics, e.g., $p$-values.

This scenario described above was addressed by \citet{Bartroff10e} who gave a procedure
that sequentially (or group sequentially) tests $k$ hypotheses while controlling the type~I familywise error  rate \citep[FWER, see][]{Hochberg87}, i.e., the probability of rejecting any true hypotheses, at a prescribed level. Their procedure requires only the existence of basic sequential tests for each data stream and makes no assumptions about the dependence between the different data streams; in particular, the error control holds when the streams are highly positively correlated, as is often the case in the application areas mentioned above. The current paper introduces a procedure to test $k$ hypotheses while simultaneously controlling both the type I and II FWERs (defined precisely below) at prescribed levels in the same general setting: No assumptions are made about the dependence between the different data streams.  We call this new procedure the \emph{sequential Holm procedure} because of its relation to Holm's \citeyearpar{Holm79} seminal fixed-sample ``step-down'' procedure which controls the FWER. Following a review of relevant previous work, we give a general formulation of the sequential Holm procedure in Section~\ref{sec:form}. In Section~\ref{sec:stats} we consider simple hypotheses and then composite hypotheses, simulation studies in Section~\ref{sec:sims}, and finally a discussion of future extensions and a summary.

\subsection{Background and Previous Work}

Separately, multiple testing and sequential testing are both quite mature fields, the latter dating back to Wald's \citeyearpar{Wald47} invention of sequential analysis following World War~II (see \citet{Siegmund85} for a summary of the major developments). Work on multiple testing dates back to classical ``multiple comparison'' procedures of \citet{Fisher32}, \citet{Scheffe53}, Tukey, and others \citep[see][]{Seber03} for testing hypotheses about parameter vectors in linear models. \citet{Holm79} proposed a general method for testing a list of null hypotheses $H^{(1)},\ldots,H^{(k)}$ that controls the type~I FWER without making any assumptions about the structure of the hypotheses or correlations between the valid, fixed-sample $p$-values $\what{p}^{(1)},\ldots,\what{p}^{(k)}$ associated with $H^{(1)},\ldots,H^{(k)}$, respectively. By relabeling if necessary, without loss of generality let $\what{p}^{(1)}\le\ldots\le \what{p}^{(k)}$. Given a prescribed FWER bound $\alpha\in(0,1)$, at stage $j=1,\ldots, k$, Holm's procedure accepts $H^{(j)},\ldots,H^{(k)}$ and terminates if  $\what{p}^{(j)}\ge\alpha/(k-j+1)$. Otherwise, $H^{(j)}$ is rejected and stage $j+1$ is commenced (provided $j<k$). \citet{Lehmann05} give a simple proof that the FWER of Holm's procedure is bounded by $\alpha$.

The intersection of multiple testing and sequential testing is less well-developed in a general setting than either individual field.  One area that has been considered is the adaptation of some classical fixed-sample tests about vector parameters, such as those mentioned above, to the sequential sampling setting, including O'Brien and Fleming's~\citeyearpar{OBrien79} sequential version of Pearson's $\chi^2$ test, and Tang et al.'s \citeyearpar{Tang89,Tang93} group sequential extensions of O'Brien's \citeyearpar{OBrien84} generalized least squares statistic. For bivariate normal populations, \citet{Jennison93} proposed a sequential test of two one-sided hypotheses about the bivariate mean vector, and  \citet{Cook94b} proposed a sequential test in  a similar setting but where one of the hypotheses is two-sided. A procedure for comparing three treatments was proposed by \citet{Siegmund93}, related to Paulson's \citeyearpar{Paulson64} earlier procedure for selecting the largest mean of $k$ normal distributions, which \citet{Bartroff10e} showed to be a special case of their more general sequential step-down method, mentioned in the previous paragraph.  Recently, \citet{Ye13} rediscovered another special case of Bartroff and Lai's \citeyearpar{Bartroff10e} procedure.

All of these procedures mentioned so far aim to explicitly control either the classical type~I error probability or  the type~I FWER. The first sequential procedures to simultaneously control both the type~I and II FWERs were introduced by \citet{De12,De12b}, who consider the $n$th set of $k$  measurements $X_n^{(1)},X_n^{(2)},\ldots,X_n^{(k)}$ from the streams~\eqref{streams} to be the measurements on the $n$th patient in a study, and patients are sampled until accept/reject decisions can be reached for each data stream. This setup is only slightly less general than the one considered here wherein individual streams can be dropped (e.g., certain measurements stopped) if they are no longer needed to reach a decision.  These procedures are compared with the proposed sequential Holm procedure in the numerical comparisons in Section~\ref{sec:sims}. The need to drop certain measurements (or endpoints), even on a patient remaining in the study, while continuing other measurements occurs frequently in practice since certain measurements may be costly or invasive. An example is the well-known Women's Health Initiative \citep[WHI,][]{Anderson04,Rossouw02}, one of the largest multiple-endpoint randomized prevention studies of its kind. The WHI dropped the endpoints designed to investigate the effect of hormone replacement therapy on cardiovascular and cancer outcomes in 2002 and 2005, respectively, but continued to follow-up participants for dementia and other cognition-related endpoints, known as the Women's Health Initiative Memory Study \citep{Espeland04,Shumaker98}.

The novelty of the procedures of \citet{De12,De12b} and those proposed herein is that they simultaneously control both the type~I and II FWERs.  For example, the procedure of \citet{Bartroff10e} allows arbitrary (i.e., non-binding) acceptances of null hypotheses while controlling the type~I FWER, hence  the relationship between these acceptances and the power of the procedure is necessarily only available by analysis on a case-by-case basis.  The current approach quantifies that relationship by, given a prescribed bound on the type~II FWER,  specifying an acceptance rule satisfying that bound.  In some applications a prescribed value of the type~II FWER may not be as readily motivated as the desired level of type~I FWER.  In this case we encourage the statistician to view the prescribed type~II FWER as a parameter to choose in order to obtain a procedure with other desirable operating characteristics, such as average sample size in the streamwise or maximum sense.

\section{General Formulation}\label{sec:form}

\subsection{Notation and Set-Up}

For simplicity of presentation we introduce the procedure in the fully-sequential setting where the possible stopping times can be any positive integer $n=1,2,\ldots$, although formulations in other settings like group-sequential and truncated settings are possible with only minor modifications. Fix the number $k\ge 2$ of data streams  and let $\bm{k}=\{1,\ldots,k\}$. Assume that there are $k$ streams~\eqref{streams} of sequentially observable data and, for each $i\in\bm{k}$, it is desired to test the null hypothesis $H^{(i)}$ versus the alternative hypothesis $G^{(i)}$ about the parameter~$\theta^{(i)}$ governing the $i$th data stream $X_1^{(i)}, X_2^{(i)},\ldots$, where $H^{(i)}$ and $G^{(i)}$ are disjoint subsets of the parameter space~$\Theta^{(i)}$ containing $\theta^{(i)}$. The individual parameters $\theta^{(i)}$ may themselves be vectors, and the \emph{global parameter} $\theta=(\theta^{(1)},\ldots,\theta^{(k)})$ is the concatenation of the individual parameters and is contained in the global parameter space $\Theta=\Theta^{(1)}\times\cdots\times \Theta^{(k)}$.

 Given $\theta\in\Theta$, let  $\mT(\theta)=\{i\in\bm{k}: \theta^{(i)}\in H^{(i)}\}$ denote the indices of the ``true'' hypotheses and $\mF(\theta)=\{i\in\bm{k}: \theta^{(i)}\in G^{(i)}\}$ the indices of the ``false'' null hypotheses. The type I and II familywise error rates, denoted \fweI$(\theta)$  and \fweII$(\theta)$, are defined as
\begin{align*}
\mbox{\fweI}(\theta)&=P_\theta(\mbox{$H^{(i)}$ is rejected for some $i\in\mT(\theta)$})\\
\mbox{\fweII}(\theta)&=P_\theta(\mbox{$H^{(i)}$ is rejected for some $i\in\mF(\theta)$}).
\end{align*} Here the notion of rejecting  (resp.\ accepting) $H^{(i)}$ is equivalent to accepting (resp.\ rejecting) $G^{(i)}$. This definition of \fweI$(\theta)$ is the same as the standard one for fixed-sample testing \citep[such as in][]{Hochberg87} and \fweII$(\theta)$ is defined analogously; the quantity  $1-\mbox{\fweII}(\theta)$ has been called ``familywise power'' by some authors \citep[e.g.,][]{Lee04}.

The building blocks of the sequential Holm procedure defined below are $k$ individual sequential test statistics $\{\Lambda^{(i)}(n)\}_{i\in\bm{k},\; n\ge 1}$, where $\Lambda^{(i)}(n)$ is the statistic for testing $H^{(i)}$ vs.\ $G^{(i)}$ based on the data $X_1^{(i)},X_2^{(i)},\ldots,X_n^{(i)}$ available  from the $i$th stream at time $n$.  Concrete examples of these test statistics are given later in this section and in Section~\ref{sec:stats} but, for now, the reader may think of $\Lambda^{(i)}(n)$ as a sequential log likelihood ratio statistic for testing $H^{(i)}$ vs.\ $G^{(i)}$, for example. Given desired \fweI$(\theta)$  and \fweII$(\theta)$ bounds $\alpha$ and  $\beta\in(0,1)$, respectively, for each data stream $i$ we assume the existence of critical values $A_s^{(i)}=A_s^{(i)}(\alpha, \beta)$ and $B_s^{(i)}=B_s^{(i)}(\alpha, \beta)$, $s\in\bm{k}$, such that
\begin{align}
P_{\theta^{(i)}}(\Lambda^{(i)}(n)\ge B_s^{(i)}\;\mbox{some $n$,}\; \Lambda^{(i)}(n')>A_1^{(i)}\;\mbox{all $n'<n$})&\le \frac{\alpha}{k-s+1}\qmq{for all}\theta^{(i)}\in H^{(i)}\label{typeI}\\
 P_{\theta^{(i)}}(\Lambda^{(i)}(n)\le A_s^{(i)}\;\mbox{some $n$,}\; \Lambda^{(i)}(n')<B_1^{(i)}\;\mbox{all $n'<n$})&\le \frac{\beta}{k-s+1}\qmq{for all}\theta^{(i)}\in G^{(i)}\label{typeII}
\end{align} for all $i, s\in\bm{k}$. We will show below that, in most cases, there are standard sequential statistics that satisfy these error bounds. Without loss of generality we assume that, for each $i\in\bm{k}$, 
\begin{equation}\label{AB.mono}
A_1^{(i)}\le A_2^{(i)}\le\ldots\le A_k^{(i)}<B_k^{(i)}\le B_{k-1}^{(i)}\le\ldots\le B_1^{(i)}.
\end{equation}
For example, if the $A_s^{(i)}$ were not non-decreasing in $s$ then they could be replaced by $\wtilde{A}_s^{(i)}=\max\{A_1^{(i)},\ldots, A_s^{(i)}\}$ for which \eqref{typeII} would still hold; similarly for $B_s^{(i)}$ and \eqref{typeI}. Note that, by \eqref{typeI}-\eqref{typeII}, the critical values $A_1^{(i)}, B_1^{(i)}$ are simply the critical values for the sequential test that samples until 
\begin{equation} \label{1hyp-cont}
A_1^{(i)}<\Lambda^{(i)}(n)<B_1^{(i)}
\end{equation} is violated, and this test has type~I and II error probabilities $\alpha/k$ and $\beta/k$, respectively.  The values $A_s^{(i)}$, $s\in\bm{k}$, are then such that the similar sequential test with critical values $A_s^{(i)}$, $B_1^{(i)}$ has type~II error probability $\beta/(k-s+1)$, and the analogous statement holds for the test with critical values $A_1^{(i)}$, $B_s^{(i)}$. 

The sequential multiple testing procedure introduced below will involve ranking the test statistics associated  with different data streams, which may be on completely different scales in general, so for each stream $i$ we introduce a \textit{standardizing function} $\vphi^{(i)}(\cdot)$ which will be applied to the statistic $\Lambda^{(i)}(n)$ before ranking. The standardizing functions  $\vphi^{(i)}$ can be any increasing functions such that $\vphi^{(i)}(A_s^{(i)})$ and $\vphi^{(i)}(B_s^{(i)})$ do not depend on $i$. For simplicity, here we take the $\vphi^{(i)}$ to be an increasing function satisfying
\begin{equation}\label{varphi}
\vphi^{(i)}(A_s^{(i)})=-(k-s+1)\qmq{and}\vphi^{(i)}(B_s^{(i)})=k-s+1\qmq{for all $s\in\bm{k}$.}
\end{equation}  For example, given $\{A_s^{(i)}, B_s^{(i)}\}_{s\in\bm{k}}$, an increasing piecewise linear function satisfying \eqref{varphi} can easily be constructed. Now let
\begin{equation}\label{stand.L}
\wtilde{\Lambda}^{(i)}(n)=\varphi^{(i)}(\Lambda^{(i)}(n))
\end{equation}
and \eqref{typeI}-\eqref{typeII} can be written as
\begin{align}
P_{\theta^{(i)}}(\wtilde{\Lambda}^{(i)}(n)\ge s\;\mbox{some $n$,}\; \wtilde{\Lambda}^{(i)}(n')>-k\;\mbox{all $n'<n$})&\le \alpha/s\qmq{for all}\theta^{(i)}\in H^{(i)}\label{typeI.stand}\\
 P_{\theta^{(i)}}(\wtilde{\Lambda}^{(i)}(n)\le -s\;\mbox{some $n$,}\; \wtilde{\Lambda}^{(i)}(n')<k\;\mbox{all $n'<n$})&\le \beta/s\qmq{for all}\theta^{(i)}\in G^{(i)}\label{typeII.stand}
\end{align} for all $i, s\in\bm{k}$.

We shall describe the sequential Holm procedure in terms of \textit{stages} of sampling, between which accept/reject decisions are made. Let $k_j\in\bm{k}$ ($j=1,2,\ldots$) denote the number of \emph{active} hypotheses (i.e., the $H^{(i)}$ which have been neither accepted nor rejected yet) at the beginning of the $j$th stage of sampling, and $n_j$ will denote the cumulative sample size of any active test statistic up to and including the $j$th stage. The total number of null hypotheses that have been rejected (resp.\ accepted) so far  at the beginning of the $j$th stage will be denoted by $r_j$ (resp.\ $a_j$). Accordingly, set $k_1=k$, $n_0=0$, $a_1=r_1=0$, and fix desired \fweI$(\theta)$  and \fweII$(\theta)$ bounds $\alpha$ and $\beta$, respectively.

\subsection{The Sequential Holm Procedure}\label{sec:Holm.proc}
The $j$th stage of sampling ($j=1,2,\ldots$) proceeds as follows, in which we let $H^{(1)},\ldots,H^{(k_j)}$ denote the active hypotheses without loss of generality.
\begin{enumerate}
\item\label{sample-step} Sample the active streams $\{X_n^{(i)}\}_{i\in\bm{k_j}, \; n>n_{j-1}}$ until $n$ equals
\begin{equation}\label{cont-samp}n_j=\inf\left\{n>n_{j-1}: -(k-a_j) < \wtilde{\Lambda}^{(i)}(n) < k-r_j \quad\mbox{is violated for some $i\in \bm{k_j}$}\right\}.\end{equation}
\item\label{step:ord} Relabel the active streams, if necessary, so that
\begin{equation}\label{orderS}
\wtilde{\Lambda}^{(1)}(n_j)\le \wtilde{\Lambda}^{(2)}(n_j)\le \ldots\le \wtilde{\Lambda}^{(k_j)}(n_j).\end{equation}
\item  
\begin{enumerate}
\item\label{acc-step} If the first inequality in \eqref{cont-samp} was violated, i.e., if $\wtilde{\Lambda}^{(i)}(n_j)\le -(k-a_j)$ for some $i\in\bm{k_j}$, then accept the $m_j\ge 1$ null hypotheses $$H^{(1)}, H^{(2)}, \ldots, H^{(m_j)},$$ where 
\begin{equation}
m_j=\min\left\{m\ge 1: \wtilde{\Lambda}^{(m+1)}(n_j)>-(k-a_j-m) \right\},\label{mj'}
\end{equation} and set $a_{j+1}=a_j+m_j$. Otherwise set $a_{j+1}=a_j$.

\item\label{rej-step} If the second inequality in \eqref{cont-samp} was violated, i.e., if $\wtilde{\Lambda}^{(i)}(n_j)\ge  k-r_j$ for some $i\in\bm{k_j}$, then reject the $m_j'\ge 1$ null hypotheses \begin{equation}\label{Hsrej}
H^{(k_j)}, H^{(k_j-1)}, \ldots, H^{(k_j-m_j'+1)},
\end{equation}
where 
\begin{equation}
m_j'=\min\left\{m\ge 1: \wtilde{\Lambda}^{(k_j-m)}(n_j)<k-r_j-m \right\},\label{mj}
\end{equation} 
and set $r_{j+1}=r_j+m_j'$. Otherwise set $r_{j+1}=r_j$.
\end{enumerate}
\item\label{stop-step} Stop if there are no remaining active hypotheses, i.e., if $a_{j+1}+r_{j+1}=k$.  Otherwise, let $k_{j+1}$ be the number of remaining active hypotheses and continue on to stage~$j+1$.
\end{enumerate}

Before giving an example of this procedure, we make some remarks about its definition.
\begin{enumerate}[(A)]
\item\label{rem:no.confl} There will never be a conflict between the acceptances in Step~\ref{acc-step} and the rejections in Step~\ref{rej-step} since if $H^{(i)}$ is accepted at stage $j$ and $i$ is the index assigned in Step~\ref{step:ord} of the $j$th stage, then $i\le m_j$, hence $i-1<m_j$ so by \eqref{mj'} we have
$$\wtilde{\Lambda}^{(i)}(n_j) \le -(k-a_j-(i-1))<0<k-r_j- (k_j-i),$$ which shows that the set in \eqref{mj} must contain the value $k_j-i$, hence $m_j'\le k_j-i$, or $i\le k_j-m_j'$. With \eqref{Hsrej}, this shows that $H^{(i)}$ could not have also been rejected. A similar argument shows that a null hypothesis that is rejected could not also be accepted at the same stage.

\item If $k=1$ then this definition becomes the sequential test \eqref{1hyp-cont} of the single null hypothesis $H^{(1)}$ versus alternative $G^{(1)}$ which has type~I and II error probabilities bounded by $\alpha$ and $\beta$, respectively.

\item Ties in \eqref{orderS} can be broken arbitrarily (at random, say) without affecting the error control proved in Theorem~\ref{thm:fwe}, below.

\item\label{rem:no.stand} If the same critical values are used for all data streams, that is, if $A_s^{(i)}=A_s^{(i')}=A_s$ and $B_s^{(i)}=B_s^{(i')}=B_s$ for all $i, i', s\in\bm{k}$, then the standardization performed in \eqref{stand.L} can be dispensed with as long as the values compared with $\wtilde{\Lambda}^{(i)}$ in the procedure's definition are replaced by the appropriate values of  $A_s$ and $B_s$. Error control still holds under these conditions, which  we prove  below as part of Theorem~\ref{thm:fwe}.

\item The critical values $A_s^{(i)}, B_s^{(i)}$ can also depend on the sample size $n$ of the test statistic being compared to them, with only notational changes in the definition of the procedure and the properties proved below.  However, to avoid overly cumbersome notation we have omitted this from the presentation. Standard group sequential stopping boundaries -- such as Pocock, O'Brien-Fleming, power family, and any others \citep[see][Chapters~2 and 4]{Jennison00} -- can be utilized for the individual test statistics in this way.
\end{enumerate}

Before stating our main result, Theorem~\ref{thm:fwe}, that this procedure controls both type I and II FWERs, we give a simplistic example to show the mechanics of the procedure. Table~\ref{tab:Bern-paths} contains three sample paths in the setting of three pairs of null and alternative hypotheses about the probability $p^{(i)}$ of success in Bernoulli data $X_n^{(i)}$, $i=1,2,3$.  Here the test statistics $\Lambda^{(i)}(n)$ are taken to be log likelihood ratios 
\begin{equation}\label{BernLLR}
\Lambda^{(i)}(n)=(2S_n^{(i)}-n)\log(.6/.4)\qmq{where}S_n^{(i)}=\sum_{j=1}^n X_j^{(i)},
\end{equation}
for testing 
\begin{equation}\label{Bern.hyp}
H^{(i)}: p^{(i)}\le .4\qmq{vs.} G^{(i)}: p^{(i)}\ge .6,\end{equation}
 $i=1,2,3$, about the success probability~$p^{(i)}$ of i.i.d.\ Bernoulli data. This choice of test statistic and calculation of the critical values given in the table's header will be explained in detail further below in Section~\ref{sec:exp.fam};  for now we merely focus on the procedure's decisions to stop or continue sampling. Per remark~(\ref{rem:no.stand}) we dispense with the standardizing functions and drop the superscript~$(i)$ from  the critical values $A_s, B_s$ with which the statistics are compared, rather than the values $\mp(k-s+1)$ from \eqref{varphi}. The values of the stopped test statistics are given in bold in Table~\ref{tab:Bern-paths}, and for clarity in this example we do not relabel the streams at each stage as in Step~\ref{step:ord} of the procedure's definition.   
 
 On sample path~1, sampling proceeds until time $n_1=7$ when $H^{(1)}$ and $H^{(2)}$ are rejected because this is the first time any of the 3 test statistics exceed $B_1$ or fall below $A_1$. In particular, $H^{(1)}$ is rejected because $\Lambda^{(1)}(7)=2.03\ge B_1=1.93$ and $H^{(2)}$ is also rejected at this time because $\Lambda^{(2)}(7)=2.03\ge B_2=1.53$ and one null hypothesis (i.e., $H^{(1)}$) has already been rejected; the fact that $\Lambda^{(2)}(7)$ also exceeds $B_1$ was not necessary for rejecting $H^{(2)}$. Next, sampling of stream 3 is continued until time $n_2=10$ when $H^{(3)}$ is accepted because its test statistic falls below $A_1=-2.43$. Similarly, on sample path~2, after rejecting $H^{(1)}$ at time $n_1=7$, $H^{(2)}$ is then rejected at time $n_2=8$ because $\Lambda^{(2)}(8)$ exceeds $B_2=1.53$ and one null hypothesis (i.e., $H^{(1)}$) has already been rejected. $H^{(3)}$ is also accepted at time $n_2=8$ for the same reason as above. On sample path~3, all three null hypotheses are rejected at time $n_1=7$ because $\Lambda^{(1)}(7)=2.03\ge B_1$, $\Lambda^{(2)}(7)=2.03\ge B_2$ and one null hypothesis (i.e., $H^{(1)}$) has already been rejected, and $\Lambda^{(3)}(7)=1.22\ge B_3$ and two null hypotheses (i.e., $H^{(1)}$ and $H^{(2)}$) have already been rejected.

\begin{table}[htdp]
\caption{Three sample paths of the sequential Holm procedure for $k=3$ hypotheses about Bernoulli data using critical values $A_1=-2.34$, $A_2=-1.94$, $A_3=-1.27$, $B_1 = 1.93$, $B_2=1.53$, $B_3=.86$. The values of the stopped sequential statistics are in bold.}
\begin{center}
\begin{tabular}{cc|cccccccccc}
Data\\
Stream&&$n=1$&2&3&4&5&6&7&8&9&10\\\hline\hline
\multicolumn{12}{c}{\textit{Sample Path 1}}\\
\multirow{2}{*}{1}&$X_n^{(1)}$&0 &1 &1 &1 &1 &1 &1 &\\
&$\Lambda^{(1)}(n)$&-.41 &.00 &.41 &.81 &1.22 &1.62 &\textbf{2.03} &\\
\multirow{2}{*}{2}&$X_n^{(2)}$&1 &0 &1 &1 &1 &1 &1 &\\
&$\Lambda^{(2)}(n)$&.41 &.00 &.41 &.81 &1.22 &1.62 &\textbf{2.03} &\\
\multirow{2}{*}{3}&$X_n^{(3)}$&0 &1 &0 &0 &1 &0 &0 &0 &0 &0\\
&$\Lambda^{(3)}(n)$&-.41 &.00 &-.41 &-.81 &-.41 &-.81 &-1.22 &-1.62 &-2.03 &\textbf{-2.43}\\\hline
\multicolumn{12}{c}{\textit{Sample Path 2}}\\
\multirow{2}{*}{1}&&0 &1 &1 &1 &1 &1 &1 &\\
&&-.41 &.00 &.41 &.81 &1.22 &1.62 &\textbf{2.03} &\\
\multirow{2}{*}{2}&&1 &0 &0 &1 &1 &1 &1 &1 &\\
&&.41 &.00 &-.41 &.00 &.41 &.81 &1.22 &\textbf{1.62} &\\
\multirow{2}{*}{3}&&0 &1 &0 &0 &0 &0 &0 &0 &\\
&&-.41 &.00 &-.41 &-.81 &-1.22 &-1.62 &-2.03 &\textbf{-2.43} &
\\\hline
\multicolumn{12}{c}{\textit{Sample Path 3}}\\
\multirow{2}{*}{1}&&1 &0 &1 &1 &1 &1 &1 &\\
&&.41 &.00 &.41 &.81 &1.22 &1.62 &\textbf{2.03} &\\
\multirow{2}{*}{2}&&1 &1 &1 &0 &1 &1 &1 &\\
&&.41 &.81 &1.22 &.81 &1.22 &1.62 &\textbf{2.03} &\\
\multirow{2}{*}{3}&&0 &1 &0 &1 &1 &1 &1 &\\
&&-.41 &.00 &-.41 &.00 &.41 &.81 &\textbf{1.22} &\\\hline
\end{tabular}
\end{center}
\label{tab:Bern-paths}
\end{table}%

Next we state the result that the sequential Holm procedure controls the type~I and II FWERs, which is proved in the appendix.
\begin{theorem}\label{thm:fwe} Fix $\alpha,\beta\in(0,1)$. If the test statistics $\Lambda^{(i)}(n)$, $i\in\bm{k}$, $n\ge 1$, and critical values $A_s^{(i)}=A_s^{(i)}(\alpha,\beta)$ and $B_s^{(i)}= B_s^{(i)}(\alpha,\beta)$, $i,s\in\bm{k}$, satisfy \eqref{typeI}-\eqref{typeII}, then the sequential Holm procedure defined above in Steps~\ref{sample-step}-\ref{stop-step} satisfies \fweI$(\theta)\le\alpha$ and \fweII$(\theta)\le\beta$ for all $\theta\in\Theta$. If $A_s^{(i)}=A_s^{(i')}=A_s$ and $B_s^{(i)}=B_s^{(i')}=B_s$ for all $i, i', s\in\bm{k}$, then this conclusion still holds if we take $\varphi^{(i)}(x)=x$ for all $i\in\bm{k}$ and replace the values $\mp(k-s+1)$ in the procedure's definition by $A_s$ and $B_s$, respectively. 
\end{theorem}

\section{Constructing Test Statistics that Satisfy \eqref{typeI}-\eqref{typeII} for Individual Data Streams  }\label{sec:stats}

Since all that is needed in the above construction of the sequential Holm procedure are sequential test statistics and critical values satisfying \eqref{typeI}-\eqref{typeII} for each data stream, in this section we show how to construct them in a few different settings and give some examples.

\subsection{Simple Hypotheses and Their Use as Surrogates for Certain Composite Hypotheses}\label{sec:simple}
In this section we show how to construct the test statistics~$\Lambda^{(i)}(n)$ and critical values $\{A_s^{(i)}, B_s^{(i)}\}_{s\in\bm{k}}$ satisfying \eqref{typeI}-\eqref{typeII} for any data stream $i$ such that $H^{(i)}$ and $G^{(i)}$ are both simple hypotheses. This setting is of interest in practice because many more complicated composite hypotheses can be reduced to simple hypotheses. In this case the test statistics $\Lambda^{(i)}(n)$ will be taken to be log-likelihood ratios because of their strong optimality properties of the resulting sequential probability ratio test (SPRT); see \citet{Chernoff72}. In order to express the likelihood ratio tests in simple form, we now make the additional assumption that each data stream $X_1^{(i)},X_2^{(i)},\ldots$ constitutes independent and identically distributed data. However, we stress that this independence assumption is limited to \emph{within} each stream so that, for example, elements of $X_1^{(i)},X_2^{(i)},\ldots$ may be correlated with (or even identical to) elements of another stream $X_1^{(i')},X_2^{(i')},\ldots$.  We represent the simple null and alternative hypotheses $H^{(i)}$ and $G^{(i)}$ by the corresponding distinct density functions $h^{(i)}$ (null) and $g^{(i)}$ (alternative) with respect to some common $\sigma$-finite  measure $\mu^{(i)}$. Formally, the parameter space~$\Theta^{(i)}$ corresponding to this data stream is the set of all densities $f$ with respect to $\mu^{(i)}$, and $H^{(i)}$ is considered \textit{true} if the true density~$f^{(i)}$ satisfies $f^{(i)}=h^{(i)}$ $\mu^{(i)}$-a.s., and is \textit{false} if $f^{(i)}=g^{(i)}$ $\mu^{(i)}$-a.s. The SPRT for testing $H^{(i)}: f^{(i)}=h^{(i)}$ vs.\ $G^{(i)}: f^{(i)}=g^{(i)}$ with type I and II error probabilities $\alpha$ and $\beta$, respectively, utilizes the simple log-likelihood ratio test statistic 
\begin{equation}\label{simpleLLR}
\Lambda^{(i)}(n)=\sum_{j=1}^n \log\left(\frac{g^{(i)}(X_{j}^{(i)})}{h^{(i)}(X_{j}^{(i)})}\right)
\end{equation} and samples sequentially until $\Lambda^{(i)}(n)\le A(\alpha,\beta)$ or $\Lambda^{(i)}(n)\ge B(\alpha,\beta)$, where the critical values $A(\alpha,\beta)$ and $B(\alpha,\beta)$ satisfy
\begin{align}
P_{h^{(i)}}(\Lambda^{(i)}(n)\ge B(\alpha,\beta)\;\mbox{some $n$,}\; \Lambda^{(i)}(n')>A(\alpha,\beta)\;\mbox{all $n'<n$})&\le \alpha\label{SPRT-typeI}\\
P_{g^{(i)}}(\Lambda^{(i)}(n)\le A(\alpha,\beta)\;\mbox{some $n$,}\; \Lambda^{(i)}(n')<B(\alpha,\beta)\;\mbox{all $n'<n$})&\le\beta.\label{SPRT-typeII}
\end{align} There are a few different options for computing $A(\alpha,\beta)$ and $B(\alpha,\beta)$ in practice. They may be computed numerically via Monte Carlo or normal approximation to the log-likelihood ratio~\eqref{simpleLLR}, but the most widely-used method is to use the simple, closed-form \emph{Wald-approximations} 
\begin{equation}\label{myAB}
A(\alpha,\beta)=\log\left(\frac{\beta}{1-\alpha}\right),\quad B(\alpha,\beta)=\log\left(\frac{1-\beta}{\alpha}\right).
\end{equation} See \citet[][Section~3.3.1]{Hoel71} or \citet{Siegmund85} for a derivation. Although, in general, the inequalities in \eqref{SPRT-typeI}-\eqref{SPRT-typeII} only hold approximately  when $A(\alpha,\beta)$ and $B(\alpha,\beta)$ are given by \eqref{myAB}, \citet{Hoel71} show that the actual type I and II error probabilities when using \eqref{myAB} can only exceed $\alpha$ or $\beta$ by a negligibly small amount in the worst case, and the difference approaches $0$ for small $\alpha$ and $\beta$, which is relevant in the present multiple testing situation where we will utilize Bonferroni-type cutdowns of $\alpha$ and $\beta$. In what follows in this section we adopt \eqref{myAB} and use these to construct the critical values $A_s^{(i)}$, $B_s^{(i)}$ of the sequential Holm procedure.  The extensive simulations performed in Section~\ref{sec:sims} show that this does not lead to any exceedances of the desired FWERs. Alternative approaches would be to compute $\{A_s^{(i)}, B_s^{(i)}\}_{s\in\bm{k}}$ via Monte Carlo, as mentioned above, or to replace \eqref{myAB} by $\log\beta$ and $\log\alpha^{-1}$, respectively, for which \eqref{SPRT-typeI}-\eqref{SPRT-typeII} always hold \citep[see][]{Hoel71} and proceed similarly, but we do not explore those options here.

The next theorem shows that, neglecting Wald's approximation, the following simple expressions~\eqref{AsBs} can be used for the critical values in the sequential Holm procedure. Specifically, we show that the left-hand-sides of \eqref{typeI}-\eqref{typeII} are equal to the right-hand-sides of \eqref{aH=aS}-\eqref{bH=bS}, and hence the inequalities in \eqref{typeI}-\eqref{typeII} hold, up to Wald's approximation.

\begin{theorem}\label{thm:simple} Suppose that, for a certain data stream~$i$, $H^{(i)}: f^{(i)}=h^{(i)}$ and $G^{(i)}: f^{(i)}=g^{(i)}$ are simple hypotheses. Let $\alpha_{Wald}^{(i)}(\alpha,\beta)$ and $\beta_{Wald}^{(i)}(\alpha,\beta)$ be the values of the probabilities on the left-hand-sides of \eqref{SPRT-typeI} and \eqref{SPRT-typeII}, respectively, with $\Lambda^{(i)}(n)$ given by \eqref{simpleLLR} and $A(\alpha,\beta)$ and $B(\alpha,\beta)$ given by the Wald approximations \eqref{myAB}. Now fix $\alpha,\beta\in(0,1)$ and for $s\in\bm{k}$ let
\begin{equation*}
\alpha_s=\alpha_s(\alpha,\beta)=\frac{(k-s+1-\beta)\alpha}{(k-s+1)(k-\beta)},\quad  \beta_s=\beta_s(\alpha,\beta)=\frac{(k-s+1-\alpha)\beta}{(k-s+1)(k-\alpha)}.
\end{equation*}
Also, let $\alpha_{Holm}^{(i)}(s)$ and $\beta_{Holm}^{(i)}(s)$ denote the left-hand-sides of \eqref{typeI} and \eqref{typeII}, respectively, with $A_s^{(i)}$, $B_s^{(i)}$ given by
\begin{equation}\label{AsBs}
A_s^{(i)}=A_s^{(i)}(\alpha,\beta)=\log\left(\frac{\beta}{(1-\alpha_s)(k-s+1)}\right),\quad B_s^{(i)}=B_s^{(i)}(\alpha,\beta)=\log\left(\frac{(1-\beta_s)(k-s+1)}{\alpha}\right).
\end{equation} 
Then, for all $s\in\bm{k}$,
\begin{align}
\alpha_{Holm}^{(i)}(s)&=\alpha_{Wald}^{(i)}(\alpha/(k-s+1),\beta_s)\qm{and}\label{aH=aS}\\
\beta_{Holm}^{(i)}(s)&=\beta_{Wald}^{(i)}(\alpha_s,\beta/(k-s+1))\label{bH=bS}
\end{align} and therefore \eqref{typeI}-\eqref{typeII} hold, up to Wald's approximation, when using the critical values~\eqref{AsBs}.
\end{theorem} 

The theorem gives simple, closed form critical values \eqref{AsBs} that can be used in lieu of Monte Carlo or other methods of calculating the $2k$ critical values~$\{A_s^{(i)}, B_s^{(i)}\}_{s\in\bm{k}}$ for a stream~$i$ whose hypotheses $H^{(i)}, G^{(i)}$ are simple. Example values of \eqref{AsBs} for $\alpha=.05$ and $\beta=.2$ are given in Table~\ref{tab:crit.vals} for $k=2,\ldots,10$.

\begin{table}[htdp]
\caption{Critical values \eqref{AsBs} of the sequential Holm procedure for simple  hypotheses,  for $\alpha=.05$, $\beta=.2$, and $k=2,\ldots,10$ to two decimal places.}
\begin{center}
\begin{tabular}{c|rrrrrrrrrr}
\multirow{2}{*}{$k$}&\multicolumn{10}{c}{$A_1, \ldots, A_{k}$}\\
&\multicolumn{10}{c}{$B_1,\ldots,B_{k}$}\\\hline\hline
\multirow{2}{*}{2}&-2.28 &-1.59&&&&&&&&\\
& 3.58 &2.89&&&&&&&&\\\hline
\multirow{2}{*}{3}&-2.69 &-2.29 &-1.60&&&&&&&\\
& 4.03 &3.62 &2.93&&&&&&&\\\hline
\multirow{2}{*}{4}&-2.98 &-2.70 &-2.29 &-1.60&&&&&&\\
& 4.33 &4.04 &3.64 &2.95&&&&&&\\\hline
\multirow{2}{*}{5}&-3.21 &-2.99 &-2.70 &-2.29 &-1.60&&&&&\\
& 4.56 &4.34 &4.05 &3.65 &2.96&&&&&\\\hline
\multirow{2}{*}{6}&-3.39 &-3.21 &-2.99 &-2.70 &-2.29 &-1.60&&&&\\
& 4.75 &4.57 &4.35 &4.06 &3.66 &2.96&&&&\\\hline
\multirow{2}{*}{7}&-3.55 &-3.39 &-3.21 &-2.99 &-2.70 &-2.30 &-1.60&&&\\
& 4.91 &4.76 &4.58 &4.35 &4.07 &3.66 &2.97&&&\\\hline
\multirow{2}{*}{8}&-3.68 &-3.55 &-3.39 &-3.21 &-2.99 &-2.70 &-2.30 &-1.60&&\\
& 5.05 &4.92 &4.76 &4.58 &4.36 &4.07 &3.66 &2.97&&\\\hline
\multirow{2}{*}{9}&-3.80 &-3.68 &-3.55 &-3.40 &-3.21 &-2.99 &-2.70 &-2.30 &-1.60&\\
& 5.17 &5.05 &4.92 &4.77 &4.58 &4.36 &4.07 &3.67 &2.97&\\\hline
\multirow{2}{*}{10}&-3.91 &-3.80 &-3.68 &-3.55 &-3.40 &-3.21 &-2.99 &-2.70 &-2.30 &-1.61\\
& 5.28 &5.17 &5.05 &4.92 &4.77 &4.59 &4.36 &4.07 &3.67 &2.98\\\hline
\end{tabular}
\end{center}
\label{tab:crit.vals}
\end{table}%

\subsubsection{Example: Exponential families}\label{sec:exp.fam}

\noindent Suppose that a certain data stream $i$  is comprised of i.i.d.\ $d$-dimensional random vectors $X_1^{(i)}, X_2^{(i)},\ldots$ from a multiparameter exponential family of densities 
\begin{equation}\label{exp.fam}
X_n^{(i)}\sim f_{\theta^{(i)}}(x)=\exp[\theta^{(i)T} x-\psi^{(i)}(\theta^{(i)})],\quad n=1,2,\ldots,
\end{equation}
 where $\theta^{(i)}$ and $x$ are $d$-vectors, $(\cdot)^T$ denotes transpose, $\psi:\mathbb{R}^d\To \mathbb{R}$ is the cumulant generating function, and it is desired to test
$$H^{(i)}: \theta^{(i)}=\eta \qmq{vs.} G^{(i)}: \theta^{(i)}=\gamma$$ for given $\eta,\gamma\in\mathbb{R}^d$. Letting $S_n^{(i)}=\sum_{j=1}^nX_j^{(i)}$, the log-likelihood ratio~\eqref{simpleLLR} in this case is
\begin{equation}\label{expLLR}
\Lambda^{(i)}(n)=(\gamma-\eta)^T S_n^{(i)}-n[\psi^{(i)}(\gamma)-\psi^{(i)}(\eta)]\end{equation} and, by Theorem~\ref{thm:simple}, the critical values \eqref{AsBs} can be used, which satisfy \eqref{typeI}-\eqref{typeII} up to Wald's approximation.

As mentioned above, many more complicated testing situations reduce to this setting.  For example, the Bernoulli example~\eqref{BernLLR} for testing \eqref{Bern.hyp} can be reduced to testing $p^{(i)}=.4$ vs.\ $p^{(i)}=.6$ by considering the worst-case error probabilities under the hypotheses \eqref{Bern.hyp}, hence \eqref{BernLLR} is given by \eqref{expLLR} with $\theta^{(i)}=\log[p^{(i)}/(1-p^{(i)})]$, $\psi^{(i)}(\theta^{(i)})=-\log(1-p^{(i)})$, $\eta=\log(.4/.6)=-\gamma$, and the critical values in Table~\ref{tab:Bern-paths} are given by \eqref{AsBs} with $\alpha=\beta=.25$, this value chosen merely to produce short sample paths for the sake of the example.

\subsection{Other Composite Hypotheses}\label{other.comp}

While many composite hypotheses can be reduced to the simple-vs.-simple situation in Section~\ref{sec:simple}, the generality of Theorem~\ref{thm:fwe} does not require this and allows any type of hypotheses to be tested as long as the corresponding sequential statistics satisfy \eqref{typeI}-\eqref{typeII}. In this section we discuss the more general case of how to proceed to apply Theorem~\ref{thm:fwe} when a certain data stream $i$ is described by a multiparameter exponential family~\eqref{exp.fam} but simple hypotheses are not appropriate.  Let 
\begin{equation*}
I(\theta^{(i)},\lambda^{(i)})=(\theta^{(i)}-\lambda^{(i)})^T\nabla\psi^{(i)}(\theta^{(i)})-[\psi^{(i)}(\theta^{(i)})-\psi^{(i)}(\lambda^{(i)})]
\end{equation*} denote the Kullback-Leibler information number for the distribution~\eqref{exp.fam}, and it is desired to test
\begin{equation}\label{comp.hyp}
H^{(i)}: u(\theta^{(i)})\le u_0\qmq{vs.} G^{(i)}: u(\theta^{(i)})\ge u_1
\end{equation}where $u(\cdot)$ is a continuously differentiable real-valued function such that 
\begin{equation*}
\mbox{for all fixed $\theta^{(i)}$, $I(\theta^{(i)},\lambda^{(i)})$ is}\left(\begin{array}{c}
\mbox{decreasing} \\
\mbox{increasing}
\end{array}\right)\mbox{in $u(\lambda^{(i)})$} \left(\begin{array}{c}
 <  \\
 > 
\end{array}\right) u(\theta^{(i)}),
\end{equation*} and $u_0<u_1$ are chosen real numbers. The family of models \eqref{exp.fam} and general form of the hypotheses \eqref{comp.hyp} contain a large number of situations frequently encountered in practice, including various two-population tests that occur frequently in randomized Phase~II and Phase~III clinical trials and other situations involving nuisance parameters; see \citet[][Chapter~4]{Bartroff13}. 

Of course there are many composite hypotheses encountered in practice which do not fit into the form \eqref{comp.hyp}, such as 
\begin{equation}\label{comp.hyp2}
\wtilde{H}^{(i)}: \theta^{(i)}=\theta_0^{(i)}\qmq{vs.} \wtilde{G}^{(i)}: \theta\ne\theta_0^{(i)}
\end{equation} for some fixed $\theta_0^{(i)}$. However, by considering true values of $\theta^{(i)}$ arbitrarily close to $\theta_0^{(i)}$, it is clear that no test of \eqref{comp.hyp2} can control the type~II error probability for all $\theta^{(i)}\in \wtilde{G}^{(i)}$ in general, and since the focus here is on tests that control both the type~I and II FWERs, one would need to restrict $\wtilde{G}^{(i)}$ in some way for that to be possible, for example by modifying $\wtilde{G}^{(i)}$ to be only the $\theta^{(i)}$ such that $||\theta^{(i)}-\theta_0^{(i)}||^2\ge \delta$ for some $\delta>0$. But this restricted form fits into the framework \eqref{comp.hyp} by choosing $u(\theta^{(i)})=||\theta^{(i)}-\theta_0^{(i)}||^2$, $u_0=0$, and $u_1=\delta$. So although it is not natural to test \eqref{comp.hyp2} in the current framework of simultaneous type~I and II FWER control, it \emph{is} natural to test \eqref{comp.hyp2} when only type~I FWER control is strictly required, and that problem has already been addressed in the sequential multiple testing setting by \citet{Bartroff10e}.

The hypotheses  \eqref{comp.hyp} can be tested with sequential generalized likelihood ratio (GLR) statistics, as follows. Letting $$\what{\theta}_n^{(i)}=(\nabla\psi^{(i)})^{-1}\left(\frac{1}{n}\sum_{j=1}^n X_{j}^{(i)}\right)$$ denote the maximum likelihood estimate (MLE) of $\theta$ based on the data from the first $n$ observations, define
\begin{align}
\Lambda_H(n)&=n\left[\inf_{\lambda:\, u(\lambda)= u_0} I(\what{\theta}_n^{(i)},\lambda)\right],\label{logGLRH}\\
\Lambda_G(n)&=n\left[\inf_{\lambda:\, u(\lambda)= u_1} I(\what{\theta}_n^{(i)},\lambda)\right],\label{logGLRG}\\
\Lambda^{(i)}(n)&=\begin{cases}
+\sqrt{2n\Lambda_H(n)},&\mbox{if $u(\what{\theta}_n^{(i)})>u_0$ and $\Lambda_H(n)\ge \Lambda_G(n)$}\label{LambdaH}\\
-\sqrt{2n\Lambda_G(n)},&\mbox{otherwise,}\end{cases} 
\end{align} where
The statistics \eqref{logGLRH} and \eqref{logGLRG} are the log-GLR statistics for testing \emph{against} $H^{(i)}$ and \emph{against} $G^{(i)}$, respectively, whose signed roots in \eqref{LambdaH} have standard normal large-$n$ limiting distribution under $u(\theta^{(i)})=u_0$ and $u_1$, respectively; see \citet[][Theorem~2]{Jennison97}, whose results further show that under group sequential sampling, the signed-root statistics have asymptotically independent increments, a fact which can be used with random walk theory to find the critical values $\{A_s^{(i)}, B_s^{(i)}\}_{s\in\bm{k}}$ for $\Lambda^{(i)}(n)$ \citep[see][Chapter~4]{Bartroff13}. However, our simulation studies have shown that under the fully-sequential sampling considered here, the small-$n$ behavior of these statistics can deviate substantially from the standard normal random walk and therefore we advocate Monte Carlo determination of the critical values $\{A_s^{(i)}, B_s^{(i)}\}_{s\in\bm{k}}$ for $\Lambda^{(i)}(n)$, which then allows their inclusion in the sequential Holm procedure. 

\subsubsection{Example: Sequential Student's $t$-tests}\label{sec:t}

As an example of the setup in this section for composite hypotheses, we consider sequential Student's $t$-tests. Suppose that the data $X_1^{(i)},X_2^{(i)},\ldots$ from a certain data stream are i.i.d.\ Normal data with  mean $\mu$ and variance $\sigma^2$, both unknown, and it is desired to test the null hypothesis~$\mu\le 0$ versus the alternative $\mu\ge \delta$, for some $\delta>0$. Formally, this is a special case of the setup \eqref{comp.hyp} by taking $\wtilde{X}_j^{(i)}=(X_j^{(i)},(X_j^{(i)})^2)^T$, $\theta^{(i)}=(\theta_1^{(i)},\theta_2^{(i)})^T=(\mu/\sigma^2,-1/(2\sigma^2))^T$, $u(\theta^{(i)})=-\theta_1^{(i)}/(2\theta_2^{(i)})=\mu$,  $u_0=0$, and $u_1=\delta$, and the parameter space and hypotheses are
 \begin{gather}
\Theta^{(i)} =\mathbb{R}\times (-\infty,0),\quad H^{(i)} =\left\{(\theta_1,\theta_2)^T\in\Theta^{(i)}: \frac{-\theta_1}{2\theta_2}\le 0\right\},\nonumber\\
\mq{and} G^{(i)} =\left\{(\theta_1,\theta_2)^T\in\Theta^{(i)}: \frac{-\theta_1}{2\theta_2}\ge\delta \right\}.\label{t.Theta}
\end{gather} By standard calculations in the exponential family \citep[see][p.~106]{Bartroff06b}, the log-GLR statistics \eqref{logGLRH}-\eqref{LambdaH} are 
\begin{align}
\Lambda_H(n)&=(n/2)\log\left[1+\left(\frac{\overline{X}_n^{(i)}}{\what{\sigma}_n}\right)^2\right],\label{t.LH}\\
\Lambda_G(n)&=(n/2)\log\left[1+\left(\frac{\overline{X}_n^{(i)}-\delta}{\what{\sigma}_n}\right)^2\right],\label{t.LG}\\
\Lambda^{(i)}(n)&=\begin{cases}
+\sqrt{2n\Lambda_H(n)},&\mbox{if $\overline{X}_n^{(i)}\ge \delta/2$}\\
-\sqrt{2n\Lambda_G(n)},&\mbox{otherwise,}\end{cases} \label{t.L}
\end{align} 
where $\overline{X}_n^{(i)}$ and $\what{\sigma}_n^2$ are the usual MLE estimates of $\mu$ and $\sigma^2$, respectively, based on $X_1^{(i)},\ldots, X_n^{(i)}$.

The next lemma verifies that there are indeed critical values for which \eqref{typeI}-\eqref{typeII} hold, and that they can be found using only the standard Normal distribution. Using this result, recursive numerical integration can be used to find the critical values for which the error bounds hold conservatively, and this is the standard method used in this setting by the many sequential and group sequential software packages that exist; see \citet[][Chapter~19]{Jennison00} and \citet[][Chapter~4.3]{Bartroff13}. 

\begin{lemma}\label{lem:t} Let $A\le 0\le B$ be arbitrary values and let $Z_1,Z_2,\ldots$ be i.i.d.\ standard Normal random variables with $T_n= \overline{Z}_n/\sqrt{\Sigma_n/n}$, where $\overline{Z}_n$ and $\Sigma_n$ are the usual unbiased estimates of mean and variance, respectively, based on $Z_1,\ldots,Z_n$.  Then, in the sequential $t$-test setup above, 
\begin{multline}\label{tI}
\sup_{\theta^{(i)}\in H^{(i)}} P_{\theta^{(i)}}(\Lambda^{(i)}(n)\ge B\;\mbox{some $n$,}\; \Lambda^{(i)}(n')>A\;\mbox{all $n'<n$})\le\\
 P(T_n\ge b_n\;\mbox{some $n$,}\; T_{n'}>a_{n'}\;\mbox{all $n'<n$})
\end{multline} and
\begin{multline}\label{tII}
\sup_{\theta^{(i)}\in G^{(i)}} P_{\theta^{(i)}}(\Lambda^{(i)}(n)\le A\;\mbox{some $n$,}\; \Lambda^{(i)}(n')<B\;\mbox{all $n'<n$})\le\\
 P(T_n\le a_n\;\mbox{some $n$,}\; T_{n'}<b_{n'}\;\mbox{all $n'<n$}),
\end{multline} where $a_n=-\sqrt{(n-1)(e^{(A/n)^2}-1)}$ and $b_n=\sqrt{(n-1)(e^{(B/n)^2}-1)}$. 

Therefore, for any fixed $\alpha,\beta\in(0,1)$, there exist critical values $\{A_s^{(i)}, B_s^{(i)}\}_{s\in\bm{k}}$ such that \eqref{typeI}-\eqref{typeII} hold and which can be computed using only the standard Normal distribution of $Z_j$.
\end{lemma} 
The lemma is proved in the Appendix. 

In the sequential $t$-test setting, some authors \citep{Lai94,Bartroff06b} have proposed bounding the standard deviation parameter $\sigma$ from above by some finite value $\overline{\sigma}<\infty$. If this is the case then the  parameter space and hypotheses~\eqref{t.Theta} can be modified in the obvious way and it follows immediately from the proof of Lemma~\ref{lem:t} in the Appendix that the derived boundaries $a_n, b_n$ in \eqref{tI}-\eqref{tII} can be further refined by replacing them by $a_n+\delta_n$ and $\max\{b_n,\delta_n/2\}$, respectively, in \eqref{tI} and by $\min\{a_n,-\delta_n/2\}$ and $b_n-\delta_n$, respectively, in \eqref{tII}, where $\delta_n=\delta\sqrt{n-1}/\overline{\sigma}$.

\subsubsection{Example: Sequential Two-Population Binomial Tests}

As another example of the setup in this section for composite hypotheses, we consider comparing two binomial populations.    This setting is widely used in randomized phase~II \citep[][Section~3.2]{Bartroff08} and phase~III \citep[][Section~3.3]{Bartroff08c} clinical trials  and has recently been used to analyze RNA sequencing data \citep[see][]{Bartroff14}.

Suppose that the data $X_1^{(i)},X_2^{(i)},\ldots$ from a certain data stream is paired Binomial data $X_n^{(i)}=(Y_{1,n}^{(i)},Y_{2,n}^{(i)})^T$ where the $Y_{j,n}^{(i)}$ ($j=1,2$) are independent $\mbox{Bin}(m_j^{(i)},p_j^{(i)})$  random variables, where the $m_j^{(i)}$ are the known ``group'' sizes and the $p_j^{(i)}$ are the unknown success probabilities. In some applications, particularly the RNA sequencing example mentioned above, the group sizes may vary sequentially (i.e., with $n$) and that situation can be handled with only minor modifications to what follows.  Hypotheses that are a special case of \eqref{comp.hyp} that can be tested are $H^{(i)}: p_1^{(i)}\le p_2^{(i)}$ vs.\ $G^{(i)}: p_1^{(i)}\ge p_2^{(i)}+\delta$, for some $\delta>0$.  In particular, 
\begin{align*}
\theta^{(i)}&=\left(\log\left(\frac{p_1^{(i)}}{1-p_1^{(i)}}\right), \log\left(\frac{p_2^{(i)}}{1-p_2^{(i)}}\right)\right)^T,\\
u(\theta^{(i)})&= 1/(e^{-\theta_1^{(i)}}+1)-1/(e^{-\theta_2^{(i)}}+1) = p_1^{(i)}-p_2^{(i)},
\end{align*} with $u_0=0$ and $u_1=\delta$. The log GLR statistic~\eqref{logGLRH} [resp.\ \eqref{logGLRG}] based on the data $X_1^{(i)},\ldots, X_n^{(i)}$ takes the form
\begin{equation*}
n\sum_{j=1}^2 \left[
  \what{p}_j^{(i)}\log\left(\frac{\what{p}_j^{(i)}}{\pi_j}\right)
  +(1- \what{p}_j^{(i)})\log\left(\frac{1-\what{p}_j^{(i)}}{1-\pi_j}\right)  \right],
\end{equation*}
where $\what{p}_j^{(i)}=\sum_{\ell=1}^n Y_{j,\ell}/(nm_j^{(i)})$ is the MLE of $p_j^{(i)}$ and $\pi_j$ is the constrained MLE of $p_j^{(i)}$ subject to $p_1^{(i)}=p_2^{(i)}$ [resp.\ $p_1^{(i)}=p_2^{(i)}+\delta$].

\section{Simulation Studies}\label{sec:sims}
In this section, we compare the sequential Holm procedure (denoted SH) with the fixed-sample Holm~\citeyearpar{Holm79} procedure (denoted FH), the sequential Bonferroni procedure (denoted SB), and the sequential intersection scheme (denoted IS) proposed by \citet{De12b}.  The SB procedure uses a SPRT on each data stream with error probability bounds $\alpha/k$ and $\beta/k$ via the Wald approximations~\eqref{myAB}. That is, for each $i\in\bm{k}$, SB samples the $i$th stream until \eqref{1hyp-cont} is violated, with $A_1^{(i)}=\log[(\beta/k)/(1-\alpha/k)]$ and $B_1^{(i)}=\log[(1-\beta/k)/(\alpha/k)]$. The three sequential procedures SH, SB, and IS are the only ones we know of that control both \fweI~and \fweII. In our studies we have chosen the commonly used values of $\alpha=.05$ and $\beta=.2$, i.e., familywise power at least 80\%. This same value of $\alpha$ is used for the fixed-sample Holm procedure as well, which does not guarantee \fweII~control at a prescribed level, so we have chosen its sample size to make its familywise power approximately the same as that of the SH procedure in order to have a meaningful comparison with the sequential procedures. Below we present two sets of simulations, the first in Table~\ref{tab:indepdendent} with independent streams of Bernoulli data, and the second in Table~\ref{tab:dependent} with dependent streams of normal data generated from a multivariate normal distribution with non-identity covariance matrix.  For each scenario considered below, \fweI, \fweII, expected total sample size $EN=E(\sum_{i=1}^k N^{(i)})$ of all the data streams where $N^{(i)}$ is the total sample size of the $i$th stream, and relative savings in sample size of SH are estimated as the result of 100,000 Monte Carlo simulated batteries of $k$ sequential tests.  In each set of simulations, the data streams and hypotheses tested are similar for each data stream; we emphasize that this is only for the sake of getting a clear picture of the procedures' performance, and this uniformity is not required in order to be able to use the procedures considered.

\subsection{Setting 1: Independent Bernoulli Data Streams}\label{sec:sims.indept}
Table~\ref{tab:indepdendent} contains the operating characteristics of the above procedures for testing $k$ hypotheses of the form $$H^{(i)}: p^{(i)}\le .4\qmq{vs.}G^{(i)}: p^{(i)}\ge .6,\quad i=1,\ldots,k,$$ about the probability $p^{(i)}$ of success in the $i$th stream of i.i.d.\ Bernoulli data; additionally, the streams were generated independently of each other.  The individual test statistics \eqref{BernLLR} were used and the SH procedure used the critical values in Table~\ref{tab:crit.vals}, as described in Section~\ref{sec:simple}. The data was generated for each data stream with $p^{(i)}=.4$ or $.6$ and the second column of Table~\ref{tab:indepdendent} gives the number of hypotheses for which $p^{(i)}=.4$. The columns labeled Savings give the percent decrease in expected sample size~$EN$ of the SH relative to each other procedure. The SH procedure has substantially smaller sample size compared to the other three, saving more than 50\% compared to FH and IS in each scenario with $k\ge 5$. Like its fixed-sample analog, the SB procedure is conservative in that its attained error rates \fweI~and \fweII~are much smaller than the prescribed levels, and the IS is similar in this regard, perhaps as a result of its stopping rule and resulting large average sample size. In fact, the IS has \fweI~and \fweII~even smaller than the SB procedure. The FH procedure has larger error rates than the SB and IS procedures, but still within the prescribed bounds, due to its step-down structure, and the SH procedure has very similar error rates to FH but with much smaller expected sample sizes.

\begin{table}[htdp]
\caption{Operating characteristics of sequential and fixed-sample multiple testing procedures for $k$ streams of independent Bernoulli data.}
\begin{center}
\resizebox{\textwidth}{!}{%
\begin{tabular}{|c|c|ccc|cccc|cccc|cccc|}
\cline{3-17}
\multicolumn{2}{c}{}&\multicolumn{3}{|c|}{SH}&\multicolumn{4}{|c|}{FH}&\multicolumn{4}{|c|}{SB}&\multicolumn{4}{|c|}{IS}\\\hline
$k$& \# of true $H^{(i)}$&\fweI&\fweII&$EN$&\fweI&\fweII&$EN$&Savings&\fweI&\fweII&$EN$&Savings&\fweI&\fweII&$EN$&Savings\\\hline
\multirow{2}*{1}&1&0.048&-&17.5&-&-&-&-&0.048&-&17.5&0.0\%&0.031&-&18.1&3.8\%\\
&0&-&0.190&24.6&-&0.194&42&41.5\%&-&0.190&24.6&0.0\%&-&0.194&28.6&14.1\%\\\hline
\multirow{3}*{2}&2&0.045&-&47.6&-&-&-&-&0.048&-&56.3&15.4\%&0.025&-&64.7&26.4\%\\
&1&0.029&0.135&63.0&0.029&0.135&126&50.0\%&0.025&0.086&66.7&5.6\%&0.021&0.120&97.2&35.2\%\\
&0&-&0.165&72.7&-&0.168&126&42.3\%&-&0.161&77.1&5.8\%&-&0.091&104.0&30.1\%\\\hline
\multirow{2}*{5}&3&0.034&0.105&216.7&0.039&0.108&485&55.3\%&0.022&0.077&230.2&5.9\%&0.010&0.044&474.9&54.4\%\\
&2&0.028&0.127&230.7&0.033&0.128&490&52.9\%&0.015&0.112&247.1&6.6\%&0.012&0.044&439.9&47.6\%\\\hline
\multirow{3}*{10}&8&0.034&0.070&479.9&0.029&0.075&1200&60.0\%&0.027&0.034&532.6&9.9\%&0.011&0.026&1144.8&58.1\%\\
&5&0.027&0.111&549.6&0.045&0.112&1240&55.7\%&0.017&0.085&587.1&6.4\%&0.008&0.023&1295.0&57.6\%\\
&2&0.016&0.130&579.4&0.033&0.132&1180&50.9\%&0.007&0.129&642.8&9.9\%&0.006&0.022&1298.2&55.4\%\\\hline
\multirow{3}*{20}&16&0.035&0.067&1129.8&0.047&0.072&2860&60.5\%&0.035&0.029&1250.8&9.7\%&0.006&0.011&3095.6&63.5\%\\
&10&0.027&0.108&1273.2&0.045&0.113&3040&58.1\%&0.022&0.073&1336.5&4.7\%&0.004&0.013&3406.0&62.6\%\\
&4&0.017&0.137&1332.6&0.035&0.138&2740&51.4\%&0.009&0.116&1421.9&6.3\%&0.004&0.015&3344.1&60.2\%\\\hline
\end{tabular}}
\end{center}
\label{tab:indepdendent}
\end{table}

\subsection{Setting 2: Correlated Normal Data Streams}\label{sec:sims.dept}
Table~\ref{tab:dependent} contains the operating characteristics of the four procedures described above  for testing $k$ hypotheses of the form $$H^{(i)}: \theta^{(i)}\le0\qmq{vs.}G^{(i)}: \theta^{(i)}\ge\delta,\quad i=1,\ldots,k,$$ for known $\delta>0$, taken here to be $1$, about the mean~$\theta^{(i)}$ of i.i.d.\ normal observations with known variance~$1$, which makes up the $i$th data stream. To investigate the performance of the procedures under dependent data streams, the streams were generated from a $k$-dimensional multivariate normal distribution with mean $\theta=(\theta^{(1)},\ldots,\theta^{(k)})$, given in the third column of Table~\ref{tab:dependent}, and four different non-identity covariance matrices $M_1, M_2, M_3$, and $M_4$, given in the Appendix, which were chosen to give a variety of different scenarios of positively and negatively correlated data streams. The interaction of these various combinations of correlations with true or false null hypotheses all show somewhat similar behavior to the case of independent data streams in the previous section, in that the SH procedure has substantially smaller expected sample size than the other three procedures in all cases, more than a 30\% reduction in most cases, and that the SH procedure has \fweI~and \fweII~much closer to the prescribed values $\alpha$ and $\beta$ than the other two sequential procedures SB and IS, and similar to the FH procedure in most cases. Because the SH procedure causes more early stopping, it is interesting to note that its error control is less conservative than the other sequential procedures even in cases when data streams with true null hypotheses are positively correlated with streams having false null hypotheses, such as the second case of the $M_1$-generated data and the third case of the $M_3$-generated data

\begin{table}[htdp]
\caption{Operating characteristics of sequential and fixed-sample multiple testing procedures for $k$ streams of correlated Normal data.}
\begin{center}
\resizebox{\textwidth}{!}{%
\begin{tabular}{|c|c|c|ccc|cccc|cccc|cccc|}
\cline{4-18}
\multicolumn{3}{c}{}&\multicolumn{3}{|c|}{SH}&\multicolumn{4}{|c|}{FH}&\multicolumn{4}{|c|}{SB}&\multicolumn{4}{|c|}{IS}\\\hline
Covariance&$k$&true $\theta$&\fweI&\fweII&$EN$&\fweI&\fweII&$EN$&Savings&\fweI&\fweII&$EN$&Savings&\fweI&\fweII&$EN$&Savings\\\hline
\multirow{3}*{$M_1$}&\multirow{3}*{2}&$(1,1)$&0.024&-&10.4&-&-&-&-&0.025&-&11.6&10.0\%&0.009&-&11.7&11.3\%\\
&&$(1,0)$&0.029&0.110&12.8&0.050&0.113&20&35.9\%&0.015&0.057&13.6&5.7\%&0.027&0.108&19.5&34.3\%\\
&&$(0,0)$&-&0.087&14.3&-&0.102&20&28.5\%&-&0.086&15.6&8.5\%&-&0.046&16.7&14.4\%\\\hline
\multirow{3}*{$M_2$}&\multirow{3}*{2}&$(1,1)$&0.029&-&10.2&-&-&-&-&0.030&-&11.6&11.9\%&0.025&-&15.1&32.5\%\\
&&$(1,0)$&0.015&0.063&13.5&0.037&0.082&22&38.8\%&0.015&0.057&13.6&0.8\%&0.009&0.016&17.9&24.8\%\\
&&$(0,0)$&-&0.114&14.4&-&0.128&20&28.1\%&-&0.113&15.6&8.0\%&-&0.103&20.3&29.3\%\\\hline
\multirow{4}*{$M_3$}&\multirow{4}*{4}&$(1,1,1,1)$&0.024&-&24.6&-&-&-&-&0.025&-&29.1&15.6\%&0.007&-&38.8&36.7\%\\
&&$(1,1,0,0)$&0.013&0.051&32.4&0.032&0.058&60&46.0\%&0.0011&0.044&33.8&4.0\%&0.001&0.003&46.9&30.9\%\\
&&$(1,0,1,0)$&0.020&0.080&32.2&0.043&0.102&56&42.5\%&0.015&0.058&33.8&4.7\%&0.010&0.021&59.6&46.0\%\\
&&$(0,0,0,0)$&-&0.089&34.1&-&0.115&48&29.1\%&-&0.090&38.4&11.4\%&-&0.029&52.6&35.3\%\\\hline
\multirow{9}*{$M_4$}&\multirow{9}*{$6$}&$(1,1,1,1,1,1)$&0.022&-&40.7&-&-&-&-&0.022&-&48.8&16.5\%&0.002&-&67.1&39.2\%\\
&&$(1,1,1,1,1,0)$&0.021&0.032&46.3&0.038&0.032&108&57.2\%&0.020&0.019&51.2&9.6\%&0.003&0.002&85.2&45.7\%\\
&&$(1,1,1,1,0,0)$&0.018&0.041&50.4&0.038&0.041&108&53.3\%&0.016&0.031&53.7&6.1\%&0.007&0.001&90.0&44.0\%\\
&&$(1,1,0,1,1,0)$&0.018&0.068&53.4&0.039&0.073&102&47.7\%&0.013&0.049&56.1&4.9\%&0.004&0.014&102.3&47.8\%\\
&&$(1,1,1,0,0,0)$&0.012&0.047&53.6&0.030&0.058&102&47.5\%&0.012&0.041&56.2&4.7\%&$<0.001$&$<0.001$&83.0&35.4\%\\
&&$(1,1,0,0,0,0)$&0.011&0.072&55.5&0.031&0.089&96&42.2\%&0.008&0.061&58.6&5.3\%&$<0.001$&0.022&97.8&43.3\%\\
&&$(1,0,0,1,0,0)$&0.016&0.074&55.3&0.045&0.092&96&42.4\%&0.010&0.059&58.7&5.7\%&0.005&0.008&104.1&46.9\%\\
&&$(1,0,0,0,0,0)$&0.008&0.077&56.3&0.040&0.099&90&37.5\%&0.005&0.071&61.1&7.9\%&0.001&0.011&97.8&42.4\%\\
&&$(0,0,0,0,0,0)$&-&0.082&55.7&-&0.087&84&33.7\%&-&0.082&63.6&12.5\%&-&0.009&89.4&37.8\%\\\hline
\end{tabular}}
\end{center}
\label{tab:dependent}
\end{table}

We have performed additional simulations in this setting but without the assumption of known variance, using the sequential $t$-test methodology in Section~\ref{sec:t}. The relative performance of the SH, FH, and SB procedures is similar to that seen in Table~\ref{tab:dependent}, with SH providing a large improvement over FH in terms of averge sample size and a moderate improvement over SB, and all with slightly larger sample sizes due to the need to estimate the additional parameter, variance.

%

\section{Discussion}\label{sec:disc}
The sequential Holm procedure proposed herein is a general method for combining individual sequential tests into a sequential multiple hypothesis testing procedure which controls both the type~I and II FWERs at prescribed levels without requiring the statistician to have any knowledge or model of the data streams' correlation structure, a desirable property that it inherits from Holm's fixed-sample procedure. In our simulations in Section~\ref{sec:sims}, the sequential Holm procedure exhibits much more efficiency in terms of smaller average total sample size than existing sequential procedures, as well as Holm's fixed-sample test. In terms of achieved FWERs, our simulations suggest that the sequential Holm procedure occupies a ``middle ground'' between existing sequential procedures, which have very conservative error rates and large average sample sizes, and the fixed-sample Holm test which achieves error rates closest to the prescribed values of all the procedures considered, but has still larger sample size and lacks the flexibility and adaptive nature of the sequential procedures.

We summarize our recommendations for using the  sequential Holm procedure in practice as follows.
\begin{itemize}
\item For data streams whose hypotheses are simple, or are composite but can be reduced to considering simple hypotheses, we recommend using the sequential log-likelihood ratio statistic~\eqref{simpleLLR} with the closed-form critical values \eqref{AsBs}.

\item For data streams with composite hypotheses of the form~\eqref{comp.hyp}, we recommend using the sequential generalized likelihood ratio statistic~\eqref{LambdaH} and determining the critical values $\{A_s^{(i)}, B_s^{(i)}\}_{s\in\bm{k}}$ to satisfy \eqref{typeI}-\eqref{typeII} by Monte Carlo. For group-sequential sampling with moderate group size the critical values can be determined by normal approximation.

\item Data streams with still other forms of hypotheses or test statistics (e.g., nonparametric) can be included in the sequential Holm procedure by determining critical values $\{A_s^{(i)}, B_s^{(i)}\}_{s\in\bm{k}}$ satisfying \eqref{typeI}-\eqref{typeII} by Monte Carlo or other methods.
\end{itemize}

As mentioned in the introduction, this subject, which lays at the intersection of sequential analysis and multiple testing, is quite young and therefore still has many interesting and fundamental unanswered questions surrounding it.  These include optimality theory for sequential multiple testing procedures, as well as calculations or estimates of their operating characteristics such as achieved FWERs and expected total  and streamwise-maximum sample size.

\section*{Acknowledgements}
The authors thank the two reviewers for suggestions that improved the paper. Bartroff's work was partially supported by National Science Foundation grants DMS-0907241 and DMS-1310127, and National Institutes of Health grant GMS-068968. The majority of Song's work was completed while a PhD student in the Department of Mathematics at the University of Southern California.

\section*{Appendix: Proofs and Details of Simulation Studies}\label{app:proofs}
\subsection*{Proofs}
\textbf{Proof of Theorem~\ref{thm:fwe}.} We fix $\theta$ and, for simplicity, omit it from the notation that follows. We first prove that \fweII$\le\beta$.  If $\mF=\emptyset$ then \fweII$=0$, so assume that $\mF\ne\emptyset$. Let 
\begin{equation}\label{Sj}
\mS_j=\left\{i\in\bm{k_j}: \;\wtilde{\Lambda}^{(i)}(n_j)\le -(k-a_j-i+1)\qm{and $H^{(i)}$ false}\right\},
\end{equation} where the indexing of the $\wtilde{\Lambda}^{(i)}$ and $H^{(i)}$ in \eqref{Sj} is with respect to the ordering in Step~\ref{step:ord} of the $j$th stage. Let $j^*$ denote the earliest stage at which a false hypothesis is accepted, taking the value $\infty$ if no such error occurs; to prove that \fweII$\le\beta$ we thus assume without loss of generality that $j^*<\infty$ with probability $1$. By our assumptions and by definition of $j^*$,  $\mS_{j^*} \ne\emptyset$ so let $i^*=\min \mS_{j^*}$. Let $\phi_{j^*}$ be the number of false hypotheses active at the beginning of stage $j^*$, and $\phi_{<j^*}=|\mF|-\phi_{j^*}$ the number of false hypotheses rejected at some stage prior to $j^*$, where $|\cdot|$ denotes set cardinality. Clearly $\phi_{<j^*}\le r_{j^*}$, and by definition of $i^*$ and $j^*$, 
$$\phi_{j^*}\le k_{j^*}-(i^*-1)=(k-a_{j^*}-r_{j^*})-(i^*-1).$$ Combining these gives
\begin{equation}\label{Tbound}
|\mF|=\phi_{j^*}+\phi_{<j^*} \le k-a_{j^*}-i^*+1.
\end{equation} 

For the remainder of the proof, let the generic index~$i\in\bm{k}$ denote the fixed, original indexing of data streams rather than an ordered index assigned in Step~\ref{step:ord} of the procedure, but let all other indices be ordered. Let $V_i$ be the event that $H^{(i)}$ is accepted at stage $j^*$ and let $i^*$ denote the index assigned to $\wtilde{\Lambda}^{(i)}(n)$ in Step~\ref{step:ord} of the $j^*$th stage. Then
\begin{equation}
\label{V.acc}
V_i\subseteq \left\{\wtilde{\Lambda}^{(i)}(n_{j^*})\le -(k-a_{j^*}-i^*+1)\right\}\subseteq \left\{\wtilde{\Lambda}^{(i)}(n_{j^*})\le -|\mF|\right\}
\end{equation} by \eqref{Tbound}. We will also show that
\begin{equation}\label{V.no.rej}
V_i\subseteq\left\{\wtilde{\Lambda}^{(i)}(n)< k\qm{for all $n<n_{j^*}$}\right\}.
\end{equation} This holds because if $\wtilde{\Lambda}^{(i)}(n)\ge k$ for some $n<n_{j^*}$, then $H^{(i)}$ would be rejected at some stage prior to $j^*$. To see this, let $W_i$ be the event on the right-hand-side of \eqref{V.no.rej}. It is clear from Step~\ref{rej-step} that on $W_i^c$, \emph{some} hypothesis would be rejected at a stage $j<j^*$ since $k\ge k-r_j$ for any value of $r_j$, and this is the rejection boundary for statistics in \eqref{cont-samp}. Let $j'<j^*$ denote the earliest stage such that $H^{(i)}$ is not rejected before stage $j'$ and
\begin{equation}\label{Lj'>B}
\wtilde{\Lambda}^{(i)}(n_{j'})\ge k.
\end{equation}
Let $k_{j'}-i'$ denote the ordered index assigned to $\wtilde{\Lambda}^{(i)}(n_{j'})$ in Step~\ref{step:ord} of the $j'$th stage. We will show that 
$$\wtilde{\Lambda}^{(k_{j'}-m)}(n_{j'})\ge k-r_{j'}-m\qmq{for all} 1\le m\le i'$$ which, by \eqref{mj}, implies that $H^{(i)}$ is rejected at stage $j'$ and finishes the proof of \eqref{V.no.rej}. For any $1\le m\le i'$, 
$$\wtilde{\Lambda}^{(k_{j'}-m)}(n_{j'})\ge \wtilde{\Lambda}^{(k_{j'}-i')}(n_{j'})\ge k \ge k-r_{j'}-m,$$
by \eqref{Lj'>B}. 

Combining \eqref{V.acc} and \eqref{V.no.rej} we have
\begin{equation*}
V_i\subseteq \left\{\wtilde{\Lambda}^{(i)}(n)\le  -|\mF|\;\mbox{some $n$,}\; \wtilde{\Lambda}^{(i)}(n')<k\;\mbox{all $n'<n$}\right\},
\end{equation*} and thus $P(V_i)\le \beta/|\mF|$ by \eqref{typeII.stand}.
Using this we have
$$\mbox{\fweII}=P\left(\bigcup_{i\in\mF} V_i\right) \le\sum_{i\in\mF}P(V_i) \le\sum_{i\in\mF}\beta/|\mF| =\beta.$$

The proof that \fweI$\le\alpha$ is similar so the details are omitted. The only thing  that could make the situation different is the possibility that a hypothesis that would have been rejected in Step~\ref{rej-step} is accepted in Step~\ref{acc-step}. However, Remark~(\ref{rem:no.confl}) guarantees that this does not happen.

For the second claim of the theorem, since $\varphi^{(i)}(\cdot)$ is strictly increasing, its inverse could be applied to $\wtilde{\Lambda}^{(i)}$ and the corresponding critical value at every place they are compared in the procedure's definition and the proof, thus replacing $\wtilde{\Lambda}^{(i)}$ by $\Lambda^{(i)}$ and $\mp(k-s+1)$ by $A_s^{(i)}, B_s^{(i)}$. If $A_s^{(i)}=A_s^{(i')}=A_s$ and $B_s^{(i)}=B_s^{(i')}=B_s$ for all $i, i', s\in\bm{k}$, then this is equivalent to using $\varphi^{(i)}(x)=x$ for all $i\in\bm{k}$. \qed

\bigskip

\noindent\textbf{Proof of Theorem \ref{thm:simple}.} First note that $\alpha_s, \beta_s\in(0,1)$ for all $s\in\bm{k}$ since
$$0<\alpha_s=\frac{k-s+1-\beta}{k-s+1}\cdot \frac{\alpha}{k-\beta}<1\cdot \frac{\alpha}{k-\beta}<\frac{1}{k-1}\le 1$$ as $k\ge 2$, and similarly for $\beta_s$.
$A_s^{(i)}$ and $B_s^{(i)}$ in \eqref{AsBs} can be written as $A(\alpha_s,\beta/(k-s+1))$ and $B(\alpha/(k-s+1),\beta_s)$, respectively, and it is simple algebra to then check that $A(\alpha/(k-s+1),\beta_s)=A_1^{(i)}$ for any $s\in\bm{k}$. Then, to verify \eqref{aH=aS},
\begin{align*}
\alpha_{Holm}^{(i)}(s)&=P_{h^{(i)}}(\Lambda^{(i)}(n)\ge B_s^{(i)}\;\mbox{some $n$,}\; \Lambda^{(i)}(n')>A_1^{(i)}\;\mbox{all $n'<n$})\\
&=P_{h^{(i)}}(\Lambda^{(i)}(n)\ge B(\alpha/(k-s+1),\beta_s)\;\mbox{some $n$,}\; \Lambda^{(i)}(n')>A(\alpha/(k-s+1),\beta_s)\;\mbox{all $n'<n$}) \\
&=\alpha_{Wald}^{(i)}(\alpha/(k-s+1),\beta_s)
\end{align*}by \eqref{SPRT-typeI}. The proof of \eqref{bH=bS} is similar.\qed

\bigskip

\noindent\textbf{Proof of Lemma \ref{lem:t}.} We will prove \eqref{tI}; the proof of \eqref{tII} is similar after replacing $X_j^{(i)}$ by $\delta-X_j^{(i)}$. Let $\wtilde{\sigma}_n^2=n\what{\sigma}_n^2/(n-1)$ and $\wtilde{T}_n=\sqrt{n}\overline{X}_n^{(i)}/\wtilde{\sigma}_n$. By \eqref{t.LH}-\eqref{t.L},
$$\left\{\Lambda^{(i)}(n) \ge B\right\}=\left\{\overline{X}_n^{(i)}\ge \delta/2\qmq{and}\sqrt{2n\Lambda_H(n)}\ge B\right\}\subseteq \left\{\frac{\overline{X}_n^{(i)}}{\what{\sigma}_n}\ge \frac{b_n}{\sqrt{n-1}}\right\} = \left\{\wtilde{T}_n\ge b_n \right\}$$
and 
\begin{multline*}
\left\{\Lambda^{(i)}(n) > A\right\}=\left\{\overline{X}_{n}^{(i)}\ge \delta/2 \right\}\cup \left\{\overline{X}_{n}^{(i)}<\delta/2\qmq{and}-\sqrt{2n\Lambda_G(n)}> A\right\}\\
= \left\{\overline{X}_{n}^{(i)}\ge \delta/2 \right\}\cup \left\{\delta+\what{\sigma}_{n}\frac{a_{n}}{\sqrt{n-1} }< \overline{X}_{n}^{(i)}<\delta/2 \right\} \subseteq \left\{ \frac{\overline{X}_{n}^{(i)}}{\what{\sigma}_{n}}> \frac{a_{n}}{\sqrt{n-1}}  \right\} =\left\{ \wtilde{T}_{n}> a_{n} \right\}. 
\end{multline*} Then for any $\theta^{(i)}=(\theta_1^{(i)},\theta_2^{(i)})^T\in H^{(i)}$, the probability on the left-hand-side of \eqref{tI} is bounded above by
\begin{equation}\label{tI.upper}
P_{(0,\theta_2^{(i)})^T}\left(\wtilde{T}_n \ge b_n\;\mbox{some $n$,}\; \wtilde{T}_{n'}>a_{n'}\;\mbox{all $n'<n$}\right)
\end{equation} and since the distribution of $\wtilde{T}_n$ under $\theta^{(i)}=(0,\theta_2^{(i)})^T$ does not depend on $\theta_2^{(i)}$, we may take $\theta_2^{(i)}=-1/2$, i.e., the standard normal distribution, and thus \eqref{tI.upper} is equal to the right-hand-side of \eqref{tI}, completing the proof.\qed

\subsection*{Details of Simulation Studies}
The four covariance matrices used in the simulations for Section~\ref{sec:sims.dept} are as the following:
\begin{align*}
M_1&=\left(\begin{array}{cc}
1&0.8\\
0.8&1\end{array}\right)\\
M_2&=\left(\begin{array}{cc}
1&-0.8\\
-0.8&1\end{array}\right)\\
M_3&=\left(\begin{array}{cccc}
1&0.8&-0.6&-0.8\\
0.8&1&-0.6&-0.8\\
-0.6&-0.6&1&0.8\\
-0.8&-0.8&0.8&1\end{array}\right)\\
M_4&=\left(\begin{array}{cccccc}
1&0.8&0.6&-0.4&-0.6&-0.8\\
0.8&1&0.8&-0.4&-0.6&-0.8\\
0.6&0.8&1&-0.4&-0.6&-0.8\\
-0.4&-0.4&-0.4&1&0.8&0.6\\
-0.6&-0.6&-0.6&0.8&1&0.8\\
-0.8&-0.8&-0.8&0.6&0.8&1\end{array}\right)
\end{align*}


\def\cprime{$'$}

\end{document}